\documentclass[english,pra,showpacs,showkeys,tightenlines,secnumarabic,11pt]{revtex4}
\usepackage[T1]{fontenc}
\usepackage[latin1]{inputenc}
\usepackage{amsmath}
\usepackage{graphicx}
\usepackage{amssymb}
\usepackage{epsfig}
\usepackage{graphics}
\usepackage[mathscr]{euscript}
\usepackage{psfrag}
\usepackage{pstricks}
\usepackage{pst-node}
\usepackage{babel}

\begin{document}
\title{Disentangling correlations in Multiple Parton Interactions}
\author{G. Calucci}
\email{giorgio.calucci@ts.infn.it}
\author{D. Treleani}
\email{daniele.treleani@ts.infn.it} \affiliation{ Dipartimento di
Fisica dell'Universit\`a di Trieste and INFN, Sezione di
Trieste,\\ Strada Costiera 11, Miramare-Grignano, I-34151 Trieste,
Italy.}
\begin{abstract}
Multiple Parton Interactions are the tool to obtain information on the correlations between partons in the hadron structure. Partons may be correlated in all degrees of freedom and all different correlation terms contribute to the cross section. The contributions due to the different parton flavors can be isolated, at least to some extent, by selecting properly the final state. In the case of high energy proton-proton collisions, the effects of correlations in the transverse coordinates and in fractional momenta are, on the contrary, unavoidably mixed in the final observables. The standard way to quantify the strength of double parton interactions is by the value of the effective cross section and a small value of the effective cross section may be originated both by the relatively short transverse distance between the pairs of partons undergoing the double interaction and by a large dispersion of the distribution in multiplicity of the multi-parton distributions. 

The aim of the present paper is to show how the effects of longitudinal and transverse correlations may be disentangled by taking into account the additional information provided by double parton interactions in high energy proton-deuteron collisions.
\end{abstract}

\pacs{11.80.La; 12.38.Bx; 13.85.Hd; 13.87.-a}

\keywords{Multiple scattering, Perturbative calculations,
Inelastic scattering, Multiple production of jets}

 \maketitle

\section{Introduction}

Multiple Parton Interactions (MPI) are expected to play an important role at the LHC\cite{HERA-LHC}\cite{Perugia}\cite{Kulesza:1999zh}\cite{Snigirev:2003cq}\cite{Acosta:2004wqa}\cite{Acosta:2006bp}\cite{Hussein:2006xr}\cite{Rogers:2008ua}
\cite{Maina:2009vx}\cite{Maina:2009sj}\cite{Domdey:2009bg}\cite{Gaunt:2009re}\cite{Berger:2009cm}\cite{Corke:2009tk}. While MPI are important to estimate the background in various channels of interest for the search of new physics\cite{Hussein:2006xr}\cite{Maina:2009vx}\cite{Maina:2009sj}, a further reason of interest is that MPI are the tool to obtain information on the multi-parton correlations in the hadron structure. The inclusive cross sections of MPI processes depend in fact linearly on the multi-parton correlations\cite{Paver:1982yp}\cite{Paver:1984ux}\cite{Rogers:2009ke}.

The experimentally accessible information is represented by the scale factors which characterize the cross section with different numbers of MPI\cite{Akesson:1986iv}\cite{Abe:1997bp}\cite{Abe:1997xk}\cite{Abazov:2009gc}. While all different correlation terms contribute to the scale factors, in order to acquire a deeper insight into the non perturbative properties of the hadron structure, one needs to disentangle the effects of the different correlation terms. 
In the case of high energy proton-proton collisions, the effects of correlations of the multi-parton distributions in transverse coordinates and in fractional momenta are all mixed in the final observables. 
The strength of double parton interactions is quantified by the value of the effective cross section. The inclusive cross section of double parton interactions in hadron-hadron collisions, $\sigma_D$, is in fact given by

\begin{eqnarray}
\sigma_D=\frac{m}{2}\frac{\sigma_A\sigma_B}{\sigma_{eff}}
\end{eqnarray}

\noindent
where $\sigma_A$ and $\sigma_B$ are the inclusive cross sections corresponding to the interactions $A$ and $B$, $m=1$ when $A$ and $B$ are identical while $m=2$ when the interactions $A$ and $B$ are distinguishable. In the simplest case the effective cross section $\sigma_{eff}$ represents the transverse interaction area where double parton interactions take place. 
A small value of  $\sigma_{eff}$ may however be originated both by the relatively short transverse distance between the pairs of partons undergoing the double interaction and by a large dispersion of the distribution in multiplicity of the multi-parton distributions. It has been pointed out that  the different effects can be separated by studying double parton interactions in collisions of hadrons with heavy nuclei\cite{Strikman:2001gz}\cite{Strikman:2010bg}. When different numbers of target nucleons are involved in the multiple process, longitudinal and transverse correlations give in fact significantly different contributions to the cross section. By separating the contributions where different numbers of target nucleons are involved in the hard interaction one would hence be able to isolate the effects of longitudinal and transverse correlations. Identifying unambiguously the final state of MPI may nevertheless represent a serious challenge in the case of heavy nuclei, due to the unbalance caused by rescatterings, the energy loss of the recoil jets etc. On the other hand dynamics is much simpler and the final state is much cleaner in the case of MPI in collisions with light nuclei, which may hence provide a good handle to approach the problem. 

The purpose of the present paper is to discuss the simplest case of MPI with a light nucleus, namely the case of double parton interactions in proton-deuteron collisions.  In our approach we will avoid any assumption on the origin of correlations, whereas our goal will rather be to enlighten the connections between the different correlation parameters and the physical observables. All effects of partonic correlations in fractional momenta and in the transverse coordinates will be worked out in detail. The comparison of the double parton scattering cross sections in $pD$ and in $pp$ collisions will finally allow a model independent separation of the contributions of longitudinal and transverse correlations.

\section{Double parton scattering in proton-deuteron interactions}
\subsection{Both nucleons in the deuteron interact with large momentum exchange}
In Fig.1 we show the discontinuity of the amplitude ${\cal A}^{(2)}$ representing the contribution to the forward proton-deuteron collision amplitude of a double parton scattering diagram where both nucleons interact with large momentum exchange. The amplitude is conveniently represented by a Feynamn graph, so that the kinematics is correct and the relevant singularities are apparent.

\begin{figure}[h]
\centering
\includegraphics[width=200mm]{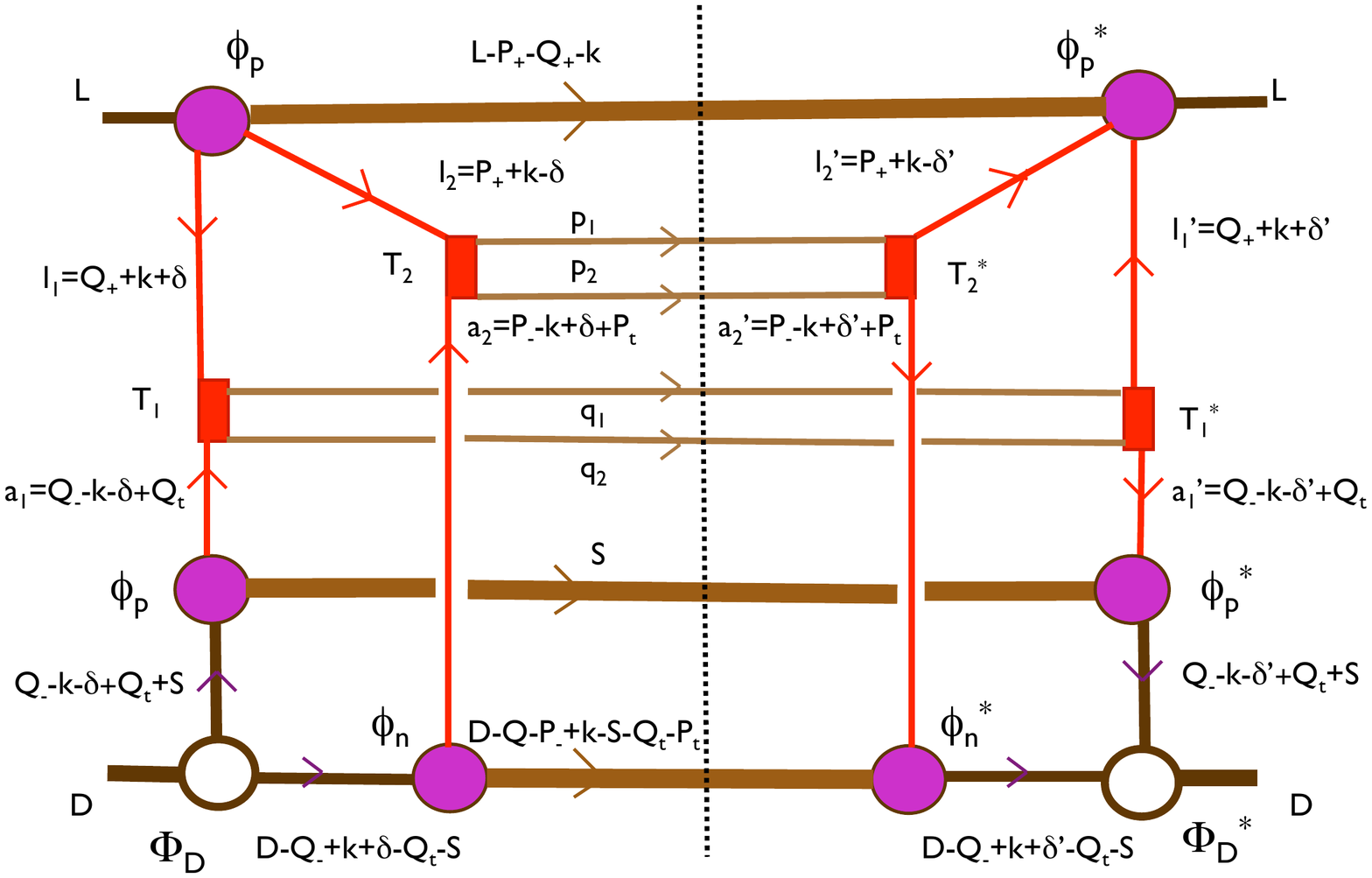}
\caption {Double parton scattering cross section in proton-deuteron interactions. Both proton and neutron interact with large momentun exchange} \label{double-two nucleons}
\end{figure}

\noindent
In the figure the two hard parton-parton interaction amplitudes are denoted $T_1$ and $T_2$, the corresponding large $p_t$ parton pairs in the final state are $p_1$, $p_2$ and $q_1$, $q_2$, with the positions $p_1+p_2=P;\;q_1+q_2=Q$. The soft vertices $\phi_p$, $\phi_n$ represent the (momentum dependent) couplings of the interacting partons with the parent proton, neutron and with the corresponding low $p_t$ final state fragments. When the vertex involves two interacting partons we put a hat, $\hat \phi$. The expressions $\Phi_D(D;N)$ are the relativistic wave function describing the nucleons with momenta $N$ and $D-N$ bound in the deuteron; their treatment is discussed in the Appendix.
The spin variables are not used: we do not consider polarized beams so we can mediate on the spin properties, the deuteron is considered a scalar state, the spinorial structure of the components disappears in the nonrelativistic limit of the relative motion.\par
The flow of momenta in the interaction is shown in the figure. The cross section is given by twice the discontinuity of the amplitude of the diagram ${\cal A}^{(2)}$ in Fig.1. The discontinuity is given by:

\begin{eqnarray}
{\rm Disc}{\cal A}^{(2)}&=&\frac{1}{(2\pi)^{21}}\int\frac{{\hat\phi}_p}{{l_1}^2 {l_2}^2}\ \frac{{\hat\phi_p}^*}{{l'_1}^2 {l'_2}^2}\ \frac{\phi_p}{a_1^2}\;
\frac{\phi_p^*}{{a'_1}^2}\;\frac{\phi_n}{{a_2}^2}\;\frac{\phi_n^*}{{a'_2}^2}\nonumber\\
&\times&T_1(l_2, a_2\to p_1,p_2)\;T_1^*(l'_2, a'_2\to p_1,p_2)\;
T_2(l_1, a_1\to q_1,q_2)\; T_2^*(l'_1,a'_1\to q_1,q_2)\nonumber\\
&\times&\frac{\Phi_D(D;N)}{[(D-N)^2-m^2][N^2-m^2]}\frac{\Phi_D^*(D;N')}{[(D-N')^2-m^2][N'^2-m^2]}\nonumber\\
&\times&\delta(L-l_1-l_2-F_3)\;\delta(L-l'_1-l'_2-F_3)\nonumber\\&\times&\delta(N-a_2-F_2)\;\delta(N'-a'_2-F_2)\;\delta(D-N-a_1-F_1)\;\delta(D-N'-a'_1-F_1)\nonumber\\&\times&\delta(l_1+a_1-Q)\:\delta(l'_1+a'_1-Q)\;\delta(l_2+a_2-P)\;\delta(l'_2+a'_2-P)\nonumber\\
& \times&\prod_{i,j}d\Omega_i \;d^4a_i d^4a'_i d^4l_i d^4l'_i d^4 F_j\delta ({F_j}^2-{M_j}^2) \;d^4N d^4N'  d^4Pd^4 Qd{M_j}^2
\label{two}
\end{eqnarray}

Note that although the expression of ${\rm Disc}{\cal A}^{(2)}$ involves the integration in $P_{\pm}$ and $Q_{\pm}$ over all the allowed range, we are interested in the regions of the phase space which correspond to a hard scattering among partons. The invariant masses of the remnants are $M_j$, $i=1,2,\;j=1,2,3$, the ten delta-functions express the four-momenta conservations in the vertices, the vectors $F_j$ represents the four momenta of the remnants of the nucleons after fragmentation, they will have masses $M_j$ certainly larger than the nucleon masses;
  $d\Omega_1$ and $d\Omega_2$ are the (two-dimensional) invariant final state phase space terms of the two hard interactions.\\
In the rest frame of the proton-nucleon centre of mass the $L_+$ and $D_-$ components grow with the c.m. energy $\sqrt{\cal S}$ while the $L_-$ and $D_+$ components decrease as  $1/\sqrt{\cal S}$. One hence has

\begin{eqnarray}
\begin{cases}
l_{1,+},\ l_{2,+},\ a_{1,-},\ a_{2,-}\lesssim{\sqrt {\cal S}}\\
l_{1,-},\ l_{2,-},\ a_{1,+},\ a_{2,+}\lesssim\frac{1}{\sqrt {\cal S}}\\
l_{1,t},\ l_{2,t},\ a_{1,t},\ a_{2,t}\lesssim\frac{1}{R}
\end{cases}
\end{eqnarray}

\noindent
being $R$ of the order of the nucleon size and independent of $\cal S $, the same holds for the $l'$ and $a'$ variables.\par
By making use of the delta-functions we may write

\begin{eqnarray}
l_1&=&(L-F_3)/2+\lambda  \qquad \qquad l'_1=(L-F_3)/2+\lambda'\cr
l_2&=&(L-F_3)/2-\lambda  \qquad \qquad l'_2=(L-F_3)/2-\lambda'\cr
a_1&=&(D-F_1-F_2)/2+\alpha  \qquad  a'_1= (D-F_1-F_2)/2+\alpha'\cr
a_2&=&(D-F_1-F_2)/2-\alpha  \qquad  a'_2 =(D-F_1-F_2)/2-\alpha'\cr
N&=& (D-F_1+F_2)/2-\alpha \qquad  N'=(D-F_1+F_2)/2-\alpha'
\end{eqnarray}

\noindent
which, besides the overall four-vectorial conservation $L+D=\sum_i F_j+P+Q$, leaves us with the integrations over $d^4\lambda,d^4\lambda',d^4\alpha,d^4\alpha',dF_j,d^4 P,d^4 Q$
 and with the remaining delta-functions\footnotemark[1]
\footnotetext[1]{apparently there is a missing
 $\delta-$function: had we chosen $\{L,D\}\neq\{L',D'\}$ we should have found a $\delta(L+D-L'-D')$, in the actual case we factor out the $\delta(0)$.} 
 
 $$4\delta(\lambda+\alpha+(P-Q)/2)\delta(\lambda'+\alpha'+(P-Q)/2)\delta(L+D-\sum_j F_j-P-Q).$$

To perform the integration over the longitudinal degrees of freedom,
 we notice that in the upper part of the diagram the $minus-$components are small while the $plus-$components are small in the lower part of the diagram. Hence the integration on $\lambda_-,\lambda'_- $, whose range of variation is of ${\cal O}(1/\sqrt {\cal S})$, affect only variables with an equally small range of variation of the $minus-$components, $i.e.$ in the upper part of the diagram, while the $minus-$components in the lower part of the diagram, whose range of variation is of ${\cal O}(\sqrt {\cal S})$, remain unaffected. The opposite property holds for all variables of type $\alpha_+,\alpha'_+$.

 At fixed values of $F_j,P,Q$ the integration in $\lambda_-,\lambda'_- $ is hence performed by disregarding the delta-functions above, since in the delta-functions all $minus-$variables, but $\lambda_-$, are large. The same attitude is taken when integrating in $\alpha_+,\alpha'_+$. The integrations in $\lambda_+,\lambda'_+,\alpha_-,\alpha'_-$ are performed, on the contrary, by using the delta-functions, where all the small terms are neglected. In so doing one obtains $\lambda_+=\lambda'_+=(Q-P)_+/2$ and $\alpha_-=\alpha'_-=(Q-P)_-/2$, which in turn implies $N_-=N'_-$, ${l_1}_+=Q_+$, ${l_2}_+=P_+$, ${a_1}_-=Q_-$, ${a_2}_-=P_-$.

 \par We proceed by defining the two-parton amplitude in the nucleon as

 $$\psi_2=\frac{1}{\sqrt 2}\int\frac{\hat\phi_p}{l_1^2 l_2^2}\frac{d\lambda_-}{2\pi i}\qquad
\psi^*_2=\frac{1}{\sqrt 2}\int\frac{\hat\phi^*_p}{{l'_1}^2 {l'_2}^2}\frac{d\lambda'_-}{2\pi i} \;.$$

The integration over $\alpha_+$ involves the vertices $\phi_p,\phi_n,\Phi_D$ and the denominator

$$[(D-N)^2-m^2]\cdot [N^2-m^2]\cdot a_1^2\cdot a_2^2$$

 In the variable $\alpha_+$ there are then four simple poles, two with positive and two with negative imaginary part. We can consider the two poles originating from the request $a_2^2=0$, or $N^2=m^2$. The first choice forces $N^2$ to be at least as large as $M_3^2$, so the nucleon is strongly off mass shell as compared with the binding energy of the deuteron. The pole $N^2=m^2$ implies a space-like
 $a_2$, which can however be near $a_2^2=0$. The integrations in $\alpha_+,\alpha'_+$ are hence approximated by keeping only the contribution of the nucleon poles.
 In this way we can define, using the relation (Eq. 3) between $N$ and $\alpha$, the covariant amplitude for finding a nucleon in the deuteron

\begin{equation}
\frac{1}{\sqrt 2} \int\frac {\Phi_D}{[(D-N)^2-m^2]\cdot [N^2-m^2]}\frac{d\alpha_+}{2\pi i}
 =\frac{1}{\sqrt 2}\frac {1}{N_-}\frac{\Phi_D}{[(D-N)^2-m^2]}\Big|_{N^2=m^2}
 =\frac{\Psi_D}{N_-}\qquad
 \end{equation}

and in strict analogy we have the twin relation:

\begin{equation}
 \frac{1}{\sqrt 2}\int\frac {\Phi_D^*}{[(D-N')^2-m^2]\cdot [N'^2-m^2]}\frac{d\alpha'_+}{2\pi i}
 =\frac{1}{\sqrt 2}\frac {1}{N'_-}\frac{\Phi_D^*}{[(D-N')^2-m^2]}\Big|_{N'^2=m^2}
 = \frac{\Psi^*_D}{N'_-}\qquad
 \end{equation}

 In this way the denominators $a_i^2$ are left as factors, in analogy with a previous definition we set

   $$\psi_{i1}=\frac{\phi_i}{a^2_i}\qquad \psi^*_{i1}=\frac{\phi_i^*}{a^2_i}\;.$$

for the one-parton amplitude one must here remember that the variables $a_i$ are, as functions of $\alpha$, subjected the condition coming from $N^2=m^2$. The dependence of the function $\psi_{i1}$ and $\psi_2$ on the invariant mass of the residual hadron fragments is understood.

\par
The integration of the longitudinal momenta of the remnants, $F_j$,  is performed by using the equivalence $d^4F \delta_+(F^2-M^2)=\frac{1}{2}dF_{\pm}/F_{\pm}\times d^2F_t$.
One may notice that ${F_3}_-$ is very small and ${F_1}_+\;{F_2}_+$ are very small too. The conditions expressed by $\delta(L+D-\sum_j F_j-P-Q)$ can hence be used to define ${F_3}_+$,
 disregarding the effect of $(F_1 +F_2)_+$, and to define $(F_1+F_2)_-$, disregarding the effect of ${F_3}_-$. In so doing from $d{F_2}_-/{F_2}_-$, $d{F_1}_-/{F_1}_-$ one gets $dK_-/[(L+D-P-Q)_-^2/4-K_-^2]$, where $K_-=(F_2- F_1)_-/2$ and it is related to $N_-$ by the relation: $K_-=(N+Q-P-D/2)_-$. At fixed $P_-,\;Q_-$ hence $dK_-=dN_-$.

 \vskip 1pc

 We can now proceed with the integration on the transverse variables, in the frame where both $L_t$, and $D_t$, are equal to zero. We take now the two-dimensional Fourier transforms (it is understood that all the variables are two-dimensional vectors).

  \begin{eqnarray}
 \psi_2&=&(2\pi)^{-2}\int\tilde\psi_2 (b_1,b_2)\exp [il_1 b_1+il_2b_2]db_1db_2\cr
 \psi_1&=&(2\pi)^{-1}\int\tilde\psi_1 (\beta_i )\exp [ia_i \beta_i] d\beta_i\cr
 \Psi&=&(2\pi)^{-1}\int\tilde\Psi (B)\exp [iN B] dB
 \end{eqnarray}

\noindent
and analogously for the complex conjugated functions, with the variables
  $b'_1,b'_2,\beta'_1,\beta'_2,B'$.\par
  The integration over the transverse-momentum variables gives the diagonal
  property $b_1=b'_1$ and so on. Moreover, as shown in Fig.2, one obtains the geometrical
  relation: $b_1-b_2=B-\beta_1-\beta_2$.

\begin{figure}[h]
\begin{center}
\includegraphics[width=120mm]{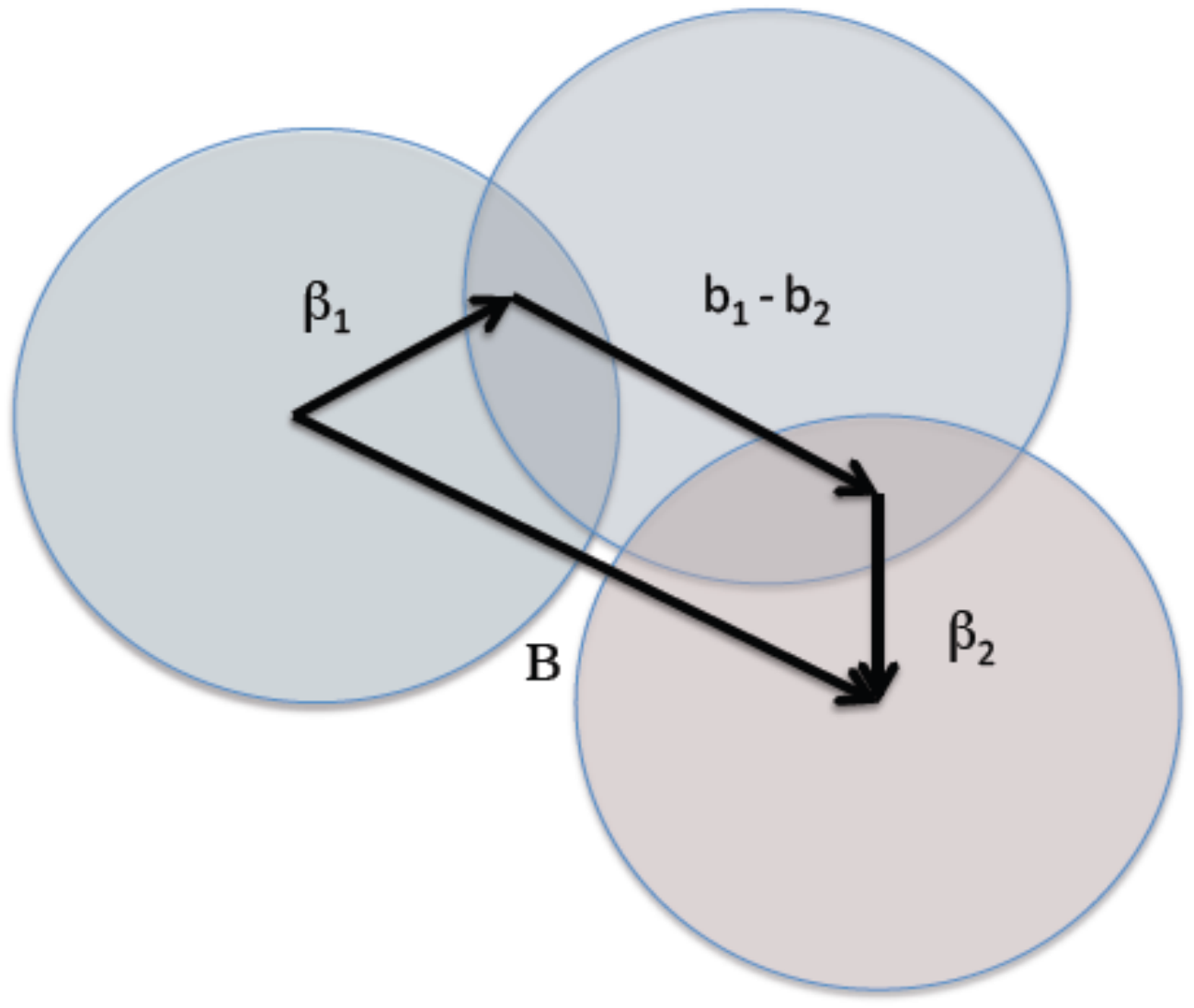}
\caption {Transverse coordinates in the double parton collision with two target nucleons involved}
\label{Overlap}
\end{center}
\end{figure}

\par
  It is useful to introduce now the fractional $plus$ or $minus$ momenta in the following way:

  $$x_1={l_1}_+/L_+,\;x_2={l_2}_+/L_+,\;x'_1=2{a_1}_-/D_-,\;x'_2=2{a_2}_/D_-$$

  We introduce also the fractional momentum of the nucleon inside the
  deuteron $ Z=2N_-/D_-$ so that the fractional momenta of the partons
  with respect to their parent nucleons are
  $z_1=x'_1/(2- Z)$, $z_2=x'_2/ Z$.

  \par
  The hard scattering amplitudes are treated as elastic scatterig between
  massless partons. Hadronization is not included. When neglecting the dependence on kinematical factors which do not grow with $\cal S$ one obtains:

\begin{eqnarray}
\frac{1}{(2\pi)^2}|T_1|^2\delta^4(P-p_1-p_2)\delta(p_1^2)\delta(p_2^2)d^4p_1d^4p_2&=&2{\cal S}x_1x_1'd\hat\sigma(x_1x_1',p_{1,t})\nonumber\\
\frac{1}{(2\pi)^2}|T_2|^2\delta^4(Q-q_1-q_2)\delta(q_1^2)\delta(q_2^2)d^4q_1d^4q_2&=&2{\cal S}x_2x_2'd\hat\sigma(x_2x_2',q_{1,t})
\end{eqnarray}

 In term of the variables $x_i,\ Z$ one has:

 \begin{eqnarray}
 dK_-&=&dN_-=(D_-/2)d Z\cr
 {F_3}_+&=&L_+ (1-x_1-x_2)\cr
 {F_1}_-&=&(D_-/2)(2- Z)(1-z_1)\cr
 {F_2}_-&=&(D_-/2) Z(1-z_2)\;.
 \end{eqnarray}

 The one-body and two-body parton densities are defined by the following integrals on the invariant mass of the residual hadron fragments:
\begin{eqnarray}
\Gamma (z,\beta)&=&\frac{1}{2(2\pi)^3}\int |\tilde\psi_M(z,\beta)|^2\frac{z}{1-z}dM^2\cr
\Gamma (x_1,x_2,b_1,b_2)&=&\frac{1}{4(2\pi)^6}\int|\tilde\psi_M(x_1,x_2,b_1,b_2)|^2\frac{2P_+Q_+}{1-x_1-x_2}dM^2
 \end{eqnarray}
The dependence on $p_{1,t} q_{1,t}$ can be transformed into an angular dependence on $\Omega_1,\;\Omega_2$ and, finally, $dP_{\pm}dQ_{\pm}={\cal S}^2 dx_1dx_2dx'_1dx'_2$. The final expression of the double parton scattering cross section where both nucleons interact $\sigma_{double}^{pD}\big|_2$ is hence given by

\begin{eqnarray}
&\sigma_{double}^{pD}\big|_2&=\frac{1}{(2\pi)^3}\int\Gamma(x_1,x_2,b_1,b_2)\frac{d\hat\sigma}{d\Omega_1}
\frac{d\hat\sigma}{d\Omega_2}\Gamma(x_1'/ Z,\beta_1)\Gamma(x_2'/(2- Z),\beta_2)\cr
&&\qquad\qquad\quad |\tilde\Psi_D( Z,B)|^2 dB\,db_1\, db_2\,d\beta_1\,d\beta_2\,\delta(B+b_1-b_2-\beta_1+\beta_2)\cr
&&\qquad\qquad\quad\big[ Z(2- Z)\big]^{-1}\,dx_1dx_2dx'_1dx'_2\,d Z\,d\Omega_1\,d\Omega_2
\end{eqnarray}

\noindent
where, as discussed in the Appendix, $\tilde\Psi_D(Z,B)$ is expressed in terms of light cone variables through the non relativistic deuteron wave function $\bar\varphi_P({\bf p}^2)$ as

\begin{eqnarray}
\tilde\Psi_D( Z,B)=\frac{1}{2\pi}\int& e^{i{\bf p_t\cdot B}}&d^2p_t\sqrt{\frac{1}{2M_D}\Bigl(\frac{ Z}{2}M_D^2+\frac{2}{ Z}m_t^2\Bigr)}\nonumber\\
&\times&\bar\varphi_P\Biggl(\frac{1}{4M_D^2}\Bigl(\frac{ Z}{2}M_D^2+\frac{2}{ Z}m_t^2\Bigr)^2-m^2\Biggr)
\end{eqnarray}

\noindent
and (as already mentioned in the Appendix) the expression has to be understood as being symmetrized for the exchange $Z\Leftrightarrow (2-Z)$.
Notice that, since

\begin{eqnarray}
\int|\tilde\Psi_D( Z,B)|^2dZd^2B=\int|\Psi_D(Z,p_t)|^2dZd^2p_t,\qquad |\Psi_D(Z,p_t)|^2=(2\pi)^3E_p|\bar\varphi_P(p)|^2
\end{eqnarray}

\noindent
and, as readily verified by using the explicit expressions of $E_p$ and $p_z$ in the Appendix, $dZ/Z=dp_z/E_p$, one has

\begin{eqnarray}
\frac{1}{(2\pi)^3}\int|\Psi_D(Z,p_t)|^2\frac{dZ}{Z}d^2p_t=\int|\bar\varphi_P(p)|^2d^3p=1
\end{eqnarray}

Given the symmetry of $\Psi_D(Z,p_t)$ for the exchange of $Z$ with $2-Z$ one moreover can easily prove the relations

\begin{eqnarray}
\frac{1}{(2\pi)^3}\int|\Psi_D(Z,p_t)|^2\frac{dZ}{Z(2-Z)}d^2p_t=\frac{1}{(2\pi)^3}\int|\Psi_D(Z,p_t)|^2\frac{dZ}{Z}d^2p_t=1
\end{eqnarray}

\noindent
which show that all densities in Eq.(11) are properly normalized.

\subsection{Only one nucleon in the deuteron interacts with large momentum exchange}

In Fig.2 we show the discontinuity of the amplitude ${\cal A}^{(1)}$ representing the contribution to the forward proton-deuteron collision amplitude of a double parton scattering diagram where one nucleon interacts with large momentum exchange and the other is a spectator.

\begin{figure}[h]
\centering
\includegraphics[width=200mm]{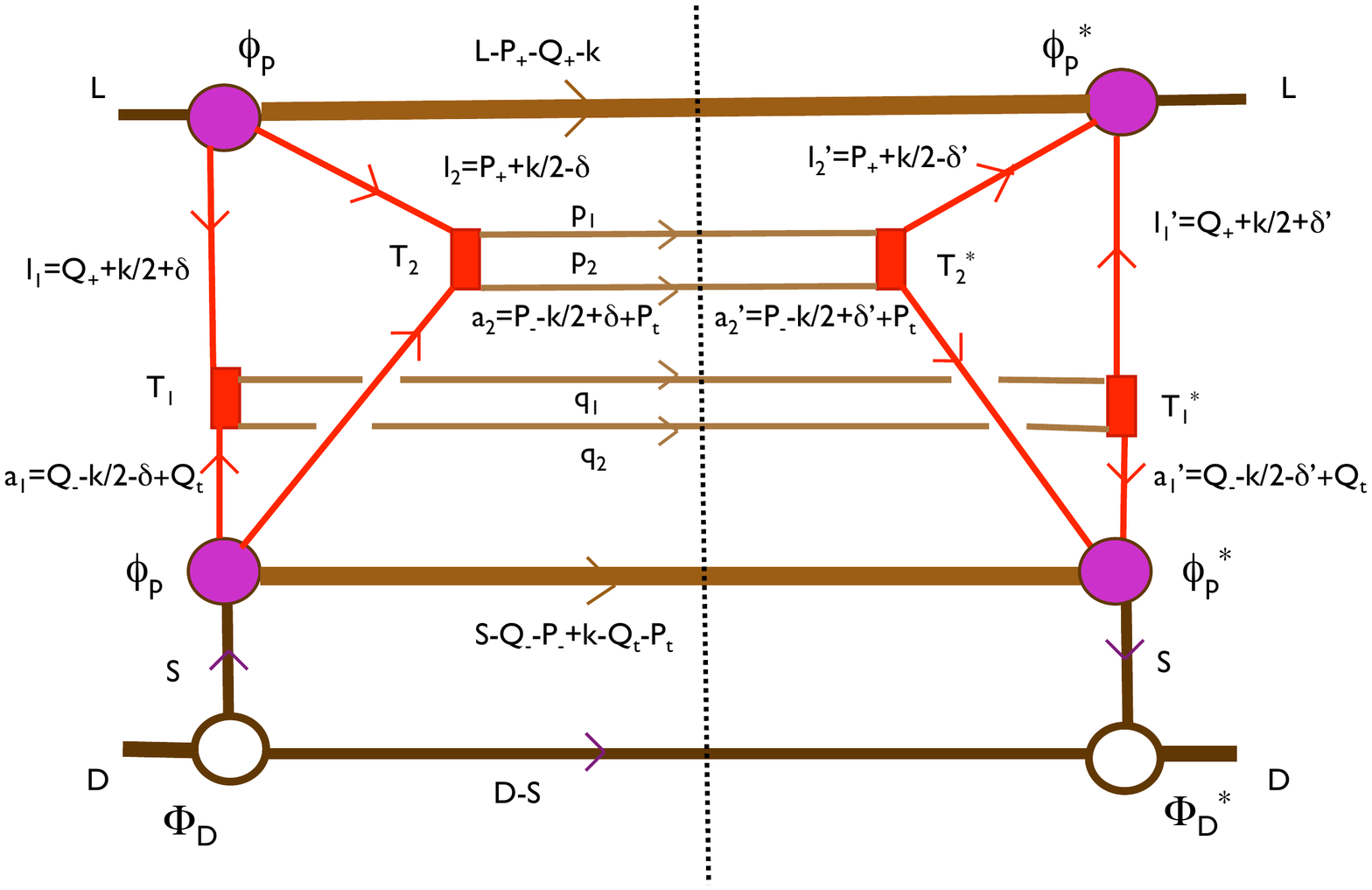}
\caption {Double parton scattering cross section in proton-deuteron interactions. The neutron is a spectator in the large momentum exchange process} \label{double-spectator nucleon}
\end{figure}

\noindent
The expression of the discontinuity is

\begin{eqnarray}
{\rm Disc}{\cal A}^{(1)}&=&\frac{1}{(2\pi)^{21}}\int\frac{\hat\phi_p}{{l_1}^2 {l_2}^2}\;\frac{\hat\phi_p^*}{{l'_1}^2{l'_2}^2}\;\frac{\hat\phi_p}{a_1^2a_2^2}\;
\frac{\phi_p^*}{{a'_1}^2{a'_2}^2}\;\nonumber\\
&\times&T_1(l_2, a_2\to p_1,p_2)\;T_1^*(l'_2, a'_2\to p_1,p_2)\;
T_2(l_1, a_1\to q_1,q_2)\; T_2^*(l'_1,a'_1\to q_1,q_2)\nonumber\\
&\times&\frac{\Phi_D}{[(D-N)^2-m^2]}\cdot\frac{\Phi_D^*}{[(D-N')^2-m^2]}
\nonumber\\
&\times&\delta(L-l_1-l_2-F_3)\;\delta(L-l'_1-l'_2-F_3)\nonumber\\&\times&
\delta(D-N-a_1-a_2-F_1)\;\delta(D-N'-a'_1-a'_2-F_1)\delta (N-F_2)\delta (N'-F_2)\nonumber\\&\times&
\delta(l_1+a_1-Q)\:\delta(l'_1+a'_1-Q)\;\delta(l_2+a_2-P)\;\delta(l'_2+a'_2-P)\nonumber\\
& \times&\prod_{i,j}d\Omega_i \;d^4a_i d^4a'_i d^4l_i d^4l'_i d^4 F_j\delta ({F_j}^2-{M_j}^2) \;d^4N d^4N'  d^4P d^4 Q
\label{two}
\end{eqnarray}

Also in this case, beyond yielding the overall four-vectorial conservation $L+D=\sum_i F_j+P+Q$, the delta-functions allow us to write

\begin{eqnarray}
l_1&=&(L-F_3)/2+\lambda  \qquad \qquad l'_1=(L-F_3)/2+\lambda'\cr
l_2&=&(L-F_3)/2-\lambda  \qquad \qquad l'_2=(L-F_3)/2-\lambda'\cr
a_1&=&(D-F_1-F_2)/2+\alpha  \qquad  a'_1= (D-F_1-F_2)/2+\alpha'\cr
a_2&=&(D-F_1-F_2)/2-\alpha  \qquad  a'_2 =(D-F_1-F_2)/2-\alpha'
\end{eqnarray}

\noindent
and obviously $\quad N=F_2=N'\;.\quad$
Multiple integrations are performed analogously to the previous paragraph. One needs to introduce the two-parton wave-function also for
the bound nucleon of the deuteron (the one defined carrying
four-momentum $S$ in Fig.3). So we define the two-parton
amplitudes of the bound nucleon as

 $${\psi_2}_p=\frac{1}{\sqrt 2}\int\frac{\hat\phi_p}{a_1^2 a_2^2}\frac{d\alpha_+}{2\pi i}\qquad
{\psi^*_2}_p=\frac{1}{\sqrt 2}\int\frac{\hat\phi^*_p}{{a'_1}^2 {a'_2}^2}\frac{d\alpha'_+}{2\pi i} \;.$$

 The spectator nucleon is set on mass shell. Notice that, in spite of that, one may still claim that final state interactions of the spectator are taken into account. The statement is supported by unitarity: If the nucleon is produced on mass shell and undergoes a final state interaction with the remnants of the other nucleon, final state interaction does not modify the inclusive cross section, since the spectator is not observed. If the nucleon is produced off mass shell, its virtuality is anyhow rather small and it may not be unreasonable to extend the unitarity relation $SS^{\dagger}=1$ to the actual kinematical domain. Thus unitarity allows us to replace the whole final state with a $cut$ nucleon line, $i.e.$ with a nucleon on mass shell.

 The different conservation relations, with respect to the case discussed in the previous paragraph, give different relations in the transverse coordinates and one obtains that the integration on the $B$ coordiante is decoupled from the others which, on the contrary, are linked by the relation $b_1-b_2=\beta_1-\beta_2$.

By following the steps and introducing the functions and the variables of the previous paragraph and keeping into account the presence of two target nucleons, one obtains the expression

\begin{eqnarray}
&\sigma_{double}^{pD}\big|_1&=\frac{2}{L_+D_-}\int|\tilde\psi_2(P_+,Q_+;b_1,b_2)_{M_1}|^22P_+P_-\frac{d\hat\sigma}{d\Omega_1}
2Q_+Q_-\frac{d\hat\sigma}{d\Omega_2}\cr&&\qquad\times|\tilde\psi_2(P_-,Q_-;\beta_1,\beta_2)_{M_2}|^2
|\tilde\Psi_D( Z,B)|^2 dB\,db_1\, db_2\,d\beta_1\,d\beta_2\,\delta(b_1-b_2-\beta_1+\beta_2)\cr
&&\qquad \times\frac{dP_+dP_-dQ_+dQ_-dN_-}{N_-(N_--Q_--P_-)(L_+-P_+-Q_+)}d\Omega_1\,d\Omega_2dM_1^2dM_2^2\frac{1}{8(2\pi)^{12}}
\end{eqnarray}

 \noindent
The double parton distribution of the bound nucleon is given by

\begin{eqnarray}
\frac{1}{Z}\Gamma (x_1'/Z,x_2'/Z;\beta_1,\beta_2)=\frac{1}{4(2\pi)^6}\int|\tilde\psi_{M_2}(P_-,Q_-;b_1,b_2)|^2\frac{2P_-Q_-}{Z-x_1'-x_2'}dM^2
 \end{eqnarray}

\noindent
and the cross section is readily expressed as:

\begin{eqnarray}
&\sigma_{double}^{pD}\big|_1&=\frac{2}{(2\pi)^3}\int\Gamma(x_1,x_2,b_1,b_2)\frac{d\hat\sigma}{d\Omega_1}
\frac{d\hat\sigma}{d\Omega_2}\Gamma(x_1'/ Z,x_2'/ Z,\beta_1,\beta_2)\cr
&&\qquad |\tilde\Psi_D( Z,B)|^2 [Z(2- Z)]^{-1}dB\,db_1\, db_2\,d\beta_1\,d\beta_2\,\delta(b_1-b_2-\beta_1+\beta_2)\cr
&&\qquad dx_1dx_2dx'_1dx'_2\,d Z\,d\Omega_1\,d\Omega_2
\end{eqnarray}

 Notice that, as an effect of the nucleon motion,  $|\tilde\Psi_D( Z,B)|^2$ is coupled to the interactions by the integration on the fractional momentum $Z$, while the integration on the transverse variable $B$ is decoupled from the other transverse variables.

\section{Some general remarks on correlations}

To the purpose of gaining some understanding of the effect of the correlations in the hadron structure on the double parton scattering cross section, in the present paragraph we discuss some basic properties of the correlation functions.

The expression $\rho(\xi)$ represents the one-body density of a set of identical particles, as a function of the variable $\xi$. The density is normalized to the average multiplicity $\langle n\rangle$

\begin{eqnarray}
\int\rho(\xi)d\xi=\langle n\rangle.
\end{eqnarray}

The two-body density $\rho(\xi_1,\xi_2)$ is analogously normalized to the second moment of the multiplicity distribution and, in general the $n$-body density is normalized to the $n$th moment of the multiplicity distribution

\begin{eqnarray}
\int\rho(\xi_1\dots\xi_n)d\xi_1\dots d\xi_n=\langle n\cdot(n-1)\dots2\cdot1\rangle
\end{eqnarray}

\noindent
Correlations are introduced in the double parton distribution as deviations from the Poissonian. One may hence express

\begin{eqnarray}
\rho(\xi_1,\xi_2)=\frac{\langle n(n-1)\rangle}{\langle n\rangle^2}\rho(\xi_1)\rho(\xi_2)+\eta(\xi_1,\xi_2)
\end{eqnarray}

\noindent
where the term $\eta(\xi_1,\xi_2)$ is such that

\begin{eqnarray}
\int\eta(\xi_1,\xi_2)d\xi_1d\xi_2=0
\end{eqnarray}

\noindent
In the uncorrelated case $\rho(\xi_1,\xi_2)=\rho(\xi_1)\rho(\xi_2)$ and one has $\langle n(n-1)\rangle=\langle n\rangle^2$, $\eta(\xi_1,\xi_2)\equiv0$. The presence of correlations thus imply $\langle n(n-1)\rangle\neq\langle n\rangle^2$ and/or $\eta(\xi_1,\xi_2)\not\equiv0$. In such a case, the distribution in multiplicity is not a Poissonian or the dependence on $\xi_1$, $\xi_2$ is not factorizable as a product $\rho(\xi_1)\rho(\xi_2)$, or both.

 For the double parton distribution one may hence write

\begin{eqnarray}
\Gamma(x_1,x_2;b_1,b_2)=K(x_1,x_2)\Gamma(x_1,b_1)\Gamma(x_2,b_2)+C(x_1,x_2;b_1,b_2)
\end{eqnarray}

\noindent
where

\begin{eqnarray}
K(x_1,x_2)&=&\frac{G(x_1,x_2)}{G(x_1)G(x_2)}\nonumber\\
G(x)&=&\int\Gamma(x,b)db\nonumber\\
G(x_1,x_2)&=&\int\Gamma(x_1,x_2;b_1,b_2)db_1db_2
\end{eqnarray}

\noindent
and

\begin{eqnarray}
\int C(x_1,x_2;b_1,b_2)db_1db_2=0
\end{eqnarray}

\noindent
All information on correlations is therefore contained in the two terms $K(x_1,x_2)$ and $C(x_1,x_2;b_1,b_2)$. When $K=1$ partons are uncorrelated after integrating in the transverse variables, while all information on the correlations on the transverse variables is contained in $C(x_1,x_2;b_1,b_2)$.

The expression of the double parton scattering cross section in proton-proton collisions may hence be written as a sum of four terms

\begin{eqnarray}
\sigma_{double}^{pp}&=&\frac{m}{2}\int\Gamma(x_1,x_2;b_1,b_2)\hat\sigma(x_1,x_1')\hat\sigma(x_2,x_2')\Gamma(x_1',x_2';b_1-B,b_2-B)\prod_{i=1}^2 dx_idx_i'db_iddB\nonumber\\
&=&\frac{m}{2}\int\Bigr\{ K(x_1,x_2)\Gamma(x_1,b_1)\Gamma(x_2,b_2)\hat\sigma(x_1,x_1')\nonumber\\
&&\qquad\qquad\times\hat\sigma(x_2,x_2')K(x_1',x_2')\Gamma(x_1',b_1-B)\Gamma(x_2',b_2-B)\nonumber\\
&&\qquad\ + K(x_1,x_2)\Gamma(x_1,b_1)\Gamma(x_2,b_2)\hat\sigma(x_1,x_1')\nonumber\\
&&\qquad\qquad\times \hat\sigma(x_2,x_2')C(x_1',x_2';b_1-B,b_2-B)\nonumber\\
&&\qquad\ + C(x_1,x_2;b_1,b_2)\hat\sigma(x_1,x_1')\nonumber\\
&&\qquad\qquad\times\hat\sigma(x_2,x_2')K(x_1',x_2')\Gamma(x_1',b_1-B)\Gamma(x_2',b_2-B)\nonumber\\
&&\qquad\ + C(x_1,x_2;b_1,b_2)\hat\sigma(x_1,x_1')\nonumber\\
&&\qquad\qquad\times\hat\sigma(x_2,x_2')C(x_1',x_2';b_1-B,b_2-B)\Bigl\}\prod_{i=1}^2 dx_idx_i'db_idB
\end{eqnarray}

\noindent
where $m=1$ in the case of two identical hard interactions and $m=2$ in the case of two different hard interactions. The hard parton-parton cross sections $\hat\sigma$ are integrated with a cutoff in momentum transfer. The four contributions to the cross section are represented in Fig.4.

\begin{figure}[h]
\begin{center}
\includegraphics[width=130mm]{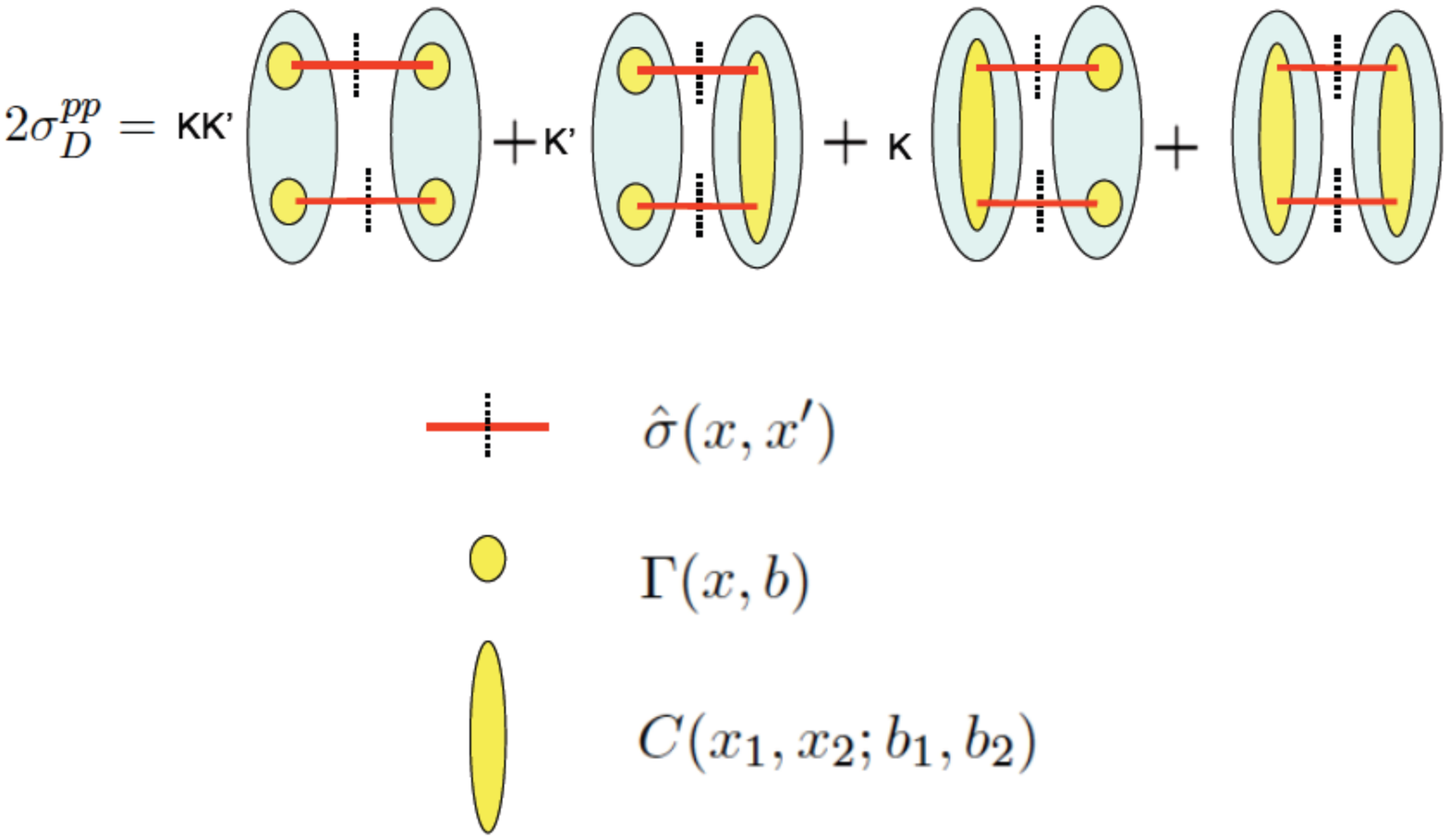}
\caption {Different contributions to the double parton scattering cross section in proton-proton collisions}
\label{Double pp}
\end{center}
\end{figure}

As evident in Fig.4 the unknown quantities related to the different source of correlations, in fractional momenta and in the transverse coordinates, cannot be disentangled by measuring the double parton scattering cross section only in proton-proton interactions. Additional independent information may be however obtained by measuring double parton collisions in proton deuteron interactions. The different contributions to the cross section are shown in Fig.5

\begin{figure}[h]
\begin{center}
\includegraphics[width=140mm]{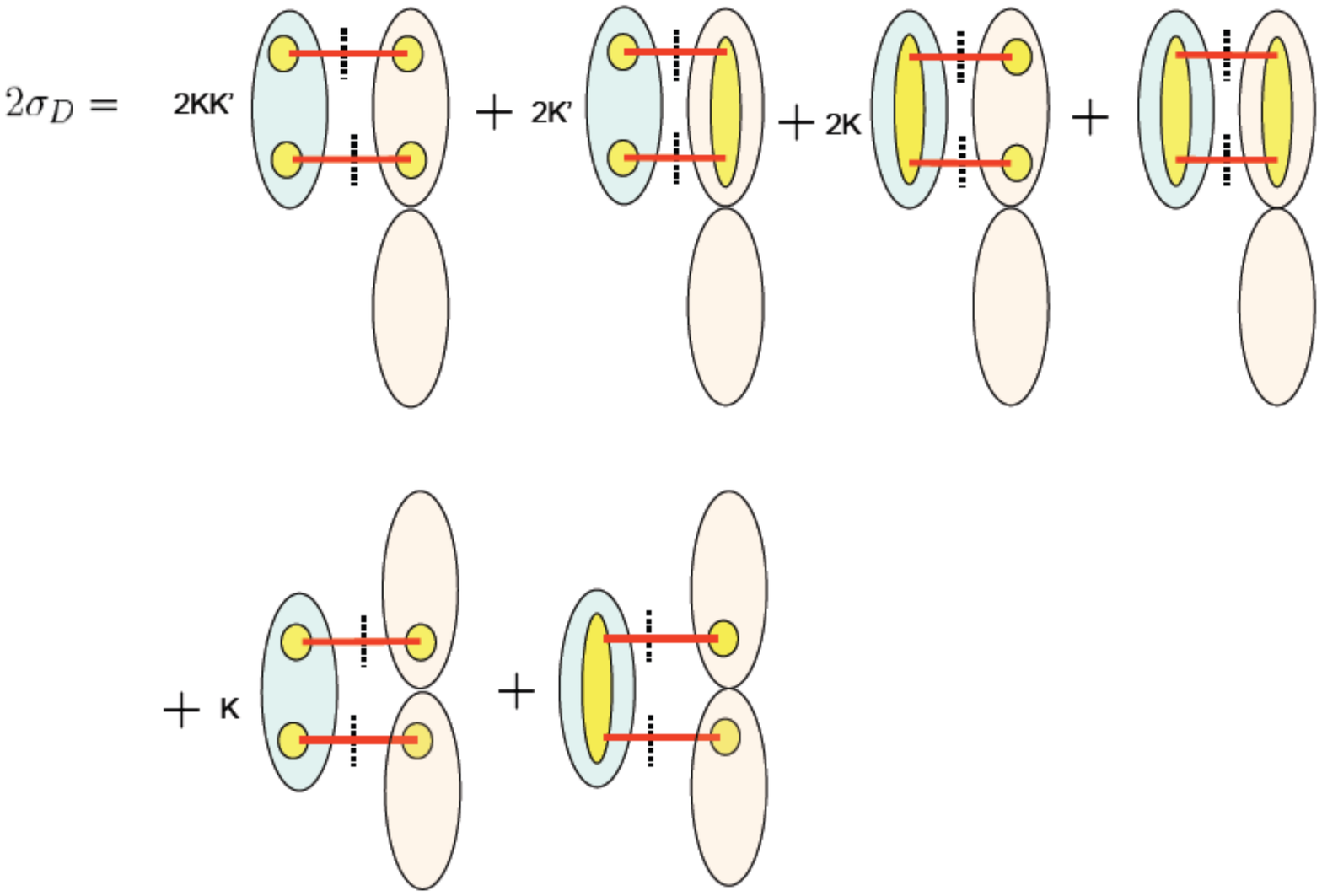}
\caption {Different contributions to the double parton scattering cross section in proton-deuteron collisions}
\label{Double pD}
\end{center}
\end{figure}

As apparent looking at the two figures 4 and 5, by comparing the double parton scattering cross sections in $pp$ and in $pD$ interactions one has the possibility to decouple the effects of longitudinal and transverse parton correlations, represented by the terms $K(x_1,x_2)$ and $C(x_1,x_2;b_1,b_2)$ in the double parton distributions.

\section{A simple estimate of the double parton cross section}

\subsection{Effects of the correlations in the transverse coordinates}

Present (still limited) evidence of double parton collisions is consistent with an effective cross section independent on fractional momenta. It wouldn't hence be inconsistent to assume $K(x_1,x_2)=K$, constant. A simplest expression of the double parton distribution where all properties previously discussed are implemented is the following factorized expressions

\begin{eqnarray}
\Gamma(x;b)&=&G(x)\frac{e^{-b^2/R^2}}{\pi R^2}\nonumber\\
\Gamma(x_1,x_2;b_1,b_2)&=&KG(x_1)G(x_2)\frac{e^{-(b_1^2+b_2^2-\lambda{\bf b}_1\cdot{\bf b}_2)/(R^2[1-\lambda^2/4])}}{(1-\lambda^2/4)(\pi R^2)^2}
\end{eqnarray}

\noindent
which satisfies

\begin{eqnarray}
\int\Gamma(x_1,x_2;b_1,b_2)db_2=KG(x_1)G(x_2)\frac{e^{-b_1^2/R^2}}{\pi R^2}=KG(x_2)\Gamma(x_1,b_1)
\end{eqnarray}

\noindent
The two parameters $K$ and $\lambda$ represent two different sources of correlations: $K\neq1$ implies a non Poissonian multiplicity distribution of partons in the relevant range of fractional momenta, while $\lambda\neq0$ implies a non trivial correlations of partons in the relative transverse coordinates. The resulting expression of the double parton scattering cross section, in the case of identical interactions in $pp$ collisions, is

\begin{eqnarray}
\sigma_{double}^{pp}&=&\frac{1}{2}\int\Gamma(x_1,x_2;b_1,b_2)\hat\sigma(x_1,x_1')\hat\sigma(x_2,x_2')\Gamma(x_1',x_2';b_1-B,b_2-B)\prod_{i=1}^2 dx_idx_i'db_idB\nonumber\\
&=&\frac{1}{2}\frac{K^2\sigma_S^2}{2\pi R^2(2-\lambda)}
\end{eqnarray}

\noindent
where $\sigma_S$ is the single scattering cross section of the QCD parton model. The effective cross section is hence related in a simple way to the correlation parameters $K$ and $\lambda$

\begin{eqnarray}
\sigma_{eff}^{pp}=\frac{2\pi R^2(2-\lambda)}{K^2}
\end{eqnarray}

\noindent
The small observed value of the effective cross section may hence imply non trivial correlations of partons in the relative transverse coordinates ($\lambda>0$) or a non Poissonian multiplicity distribution of partons at small fractional momenta ($K>1$) or both. This simplest model also shows that the two sources of correlation cannot be disentangled by measuring double parton collisions in $pp$ interactions only. An independent constraint is however provided by measuring double parton collisions in proton-nucleus interactions.
The double parton cross section with a spectator nucleon in fact is given by

\begin{eqnarray}
\sigma_{double}^{pD}\big|_1&=&\frac{2}{(2\pi)^3}\int\Gamma(x_1,x_2,b_1,b_2)\frac{d\hat\sigma}{d\Omega_1}
\frac{d\hat\sigma}{d\Omega_2}\Gamma(x_1'/ Z,x_2'/ Z,\beta_1,\beta_2)\cr
&&\qquad\qquad |\tilde\Psi_D( Z,B)|^2 [Z(2- Z)]^{-1}dB\,db_1\, db_2\,d\beta_1\,d\beta_2\,\delta(b_1-b_2-\beta_1+\beta_2)\cr
&&\qquad\qquad dx_1dx_2dx'_1dx'_2\,d Z\,d\Omega_1\,d\Omega_2\nonumber\\
&=&\frac{2K^2}{2\pi R^2(2-\lambda)}\int \sigma_S(x_1x_1'/Z)
\sigma_S(x_2x_2'/Z)|\tilde\Psi_D(Z,\beta)|^2 [(2\pi)^3Z(2- Z)]^{-1}d\beta\ dZ\nonumber\\
&=&2\int \sigma_{double}^{pp}\bigl(x_1,x'_1/Z,x_2,x_2'/Z\bigr)|\tilde\Psi_D(Z,\beta)|^2 [(2\pi)^3Z(2- Z)]^{-1}d\beta\ dZ\nonumber\\
&\simeq&2 \sigma_{double}^{pp}(x_1,x'_1,x_2,x_2')
\end{eqnarray}

\noindent
while the double parton cross section where both nucleons take part to the double hard interaction is given by

\begin{eqnarray}
&\sigma_{double}^{pD}\big|_2&=\int\Gamma(x_1,x_2,b_1,b_2)\frac{d\hat\sigma}{d\Omega_1}
\frac{d\hat\sigma}{d\Omega_2}\Gamma(x_1'/ Z,\beta_1)\Gamma(x_2'/(2- Z),\beta_2)\cr
&&\qquad\quad |\tilde\Psi_D( Z,B)|^2 dB\,db_1\, db_2\,d\beta_1\,d\beta_2\,\delta(B+b_1-b_2-\beta_1+\beta_2)\cr
&&\qquad\quad\big[ (2\pi)^3Z(2- Z)\big]^{-1}\,dx_1dx_2dx'_1dx'_2\,d Z\,d\Omega_1\,d\Omega_2\nonumber\\
&=&K\int \sigma_S\bigl(x_1x_1'/Z\bigr)
\sigma_S\bigl(x_2x_2'/(2-Z)\bigr)\nonumber\\
&&\qquad\times g\bigl(\beta, R^2(4-\lambda)\bigr)|\tilde\Psi_D(Z,\beta)|^2\big[ (2\pi)^3Z(2- Z)\big]^{-1} d\beta dZ\nonumber\\
&\simeq& K\int \sigma_S\bigl(x_1x_1'/Z\bigr)
\sigma_S\bigl(x_2x_2'/(2-Z)\bigr)|\tilde\Psi_D(Z,0)|^2 \big[ (2\pi)^3Z(2- Z)\big]^{-1}dZ\nonumber\\
&=& K\sigma_{eff}\int \sigma_{double}^{pp}\bigl(x_1,x_1'/Z,x_2,x_2'/(2-Z)\bigr)|\tilde\Psi_D(Z,0)|^2 \big[ (2\pi)^3Z(2- Z)\big]^{-1}dZ\nonumber\\
&\simeq& \sigma_{double}^{pp}(x_1,x_1',x_2,x_2')\frac{K\sigma_{eff}}{\pi R_D^2}
\end{eqnarray}

\noindent
where the function $g\bigl(\beta, R^2(4-\lambda)\bigr)$ is a Gaussian normalized to one, with argument $\beta$ and radius $[R^2(4-\lambda)]^{1/2}$. The motion of the nucleons inside the deuteron is slow, their fractional longitudinal momentum $Z$ is approximately $1$, so the nucleons distribution $D(Z,B)\equiv|\tilde\Psi_D( Z,B)|^2\big[ (2\pi)^3Z(2- Z)\big]^{-1}$ has been treated as a function $\delta(1-Z)$.  

The term $\sigma_{double}^{pD}\big|_2$ represents the contribution to the $pD$ cross section with additional information on the correlations. By comparing $\sigma_{double}^{pD}\big|_2$ with the double parton scattering cross section in $pp$ interactions one may have an indication on the value of $K$ which, measured at different values of $x_1,\ x_2$, allows to access the information on longitudinal correlations in a model independent way.
As an order of magnitude estimate the contribution of the term $\sigma_{double}^{pD}\big|_2$ to the proton deuteron double parton scattering cross section is of order $\sigma_{eff}/(2\pi R_D^2)\simeq 5\%$.

\subsection{Smearing with the Deuteron wave function along the longitudinal direction}
In the previous subsection the nucleons distribution $D(Z,B)$ has been treated as a function $\delta(1-Z)$ when integrating on $Z$. One may estimate the effect of the finite width of the deuteron by expanding the integrand around 1 as follows

\begin{eqnarray}
F(x)\to\int F(x/Z)D(Z)dZ,\quad F(x/Z)=F(x)+\frac{\partial F}{\partial Z}\Big|_{Z=1}(Z-1)+\frac{\partial^2 F}{\partial Z^2}\Big|_{Z=1}\frac{(Z-1)^2}{2}\dots
\end{eqnarray}

\noindent
Given the symmetry of $D(Z)$ around 1, the leading correction to the $\delta$-function contribution is due to the second derivative term, which is hence multiplied by $\frac{1}{2}\int(Z-1)^2D(Z)dZ$. By expanding the expression of ${\bf p}^2$ in the appendix near $Z=1$ one obtains

\begin{eqnarray}
{\bf p}^2={\bf p}_t^2+\frac{M_D^2}{4}(Z-1)^2+\dots
\end{eqnarray}

In the Deuteron wave function $(Z-1)^2$ is hence multiplied by $R^2M_D^2/4$, with $R$ a length of the order of the radius of the Deuteron. One may hence estimate

\begin{eqnarray}
\frac{1}{2}\int(Z-1)^2D(Z)dZ\approx\frac{2}{R^2M_D^2}\approx0.5\%
\end{eqnarray}

\noindent
The effect is hence roughly one order of magnitude smaller as compared to the correction due to the transverse structure of the Deuteron.

\section{Conclusions}

The inclusive cross sections of multiple parton interactions depend linearly on the multiparton correlations of the hadron structure. Partons may however be correlated in all their different degrees of freedom, while all different correlation terms contribute to the cross section.

Whereas correlations in the flavor indices may be studied by selecting properly the final states, in this paper we have focused on the problem of disentangling the contributions of the correlations in the transverse parton coordinates from the correlations in the distribution in multiplicity of the multi-parton distributions. In the case of $pp$ interactions the two contributions are unavoidably mixed. The double parton scattering cross section in $pD$ collisions can however provide additional information, which may be used to decouple the two terms.

In the present paper we have worked out in detail the two different contributions to the double parton scattering cross section in $pD$ interactions, corresponding to the different possibilities of having 1) just a single nucleon or 2) both nucleons taking part to the hard process. In the first instance the integration on the relative transverse coordinate of the two nucleons is decoupled from the other transverse variables and the scale factor in the corresponding cross section, $\sigma_{double}^{pD}\big|_1$, is (apart from minor correction terms) the same which one finds in nucleon-nucleon collisions, namely $\sigma_{eff}$. As a consequence $\sigma_{double}^{pD}\big|_1$ is twice as big as the double parton scattering cross section in $pp$ collisions. In the second instance the scale factor, in the corresponding cross section $\sigma_{double}^{pD}\big|_2$, is mainly provided by the transverse size of the deuteron. As a consequence the value of $\sigma_{double}^{pD}\big|_2$ is very sensitive to the actual value of the correlation term $K(x_1,x_2)$, which measures how much the multi-parton distribution in multiplicity is different from a Poissonian, at given values of fractional momenta.

The first term, $\sigma_{double}^{pD}\big|_1$, is the term usually taken into account when considering large $p_t$ processes in hadron-nucleus collisions. The second term, $\sigma_{double}^{pD}\big|_2$, represents a positive (namely a antishadowing) correction. A rough quantitative estimate indicates that $\sigma_{double}^{pD}\big|_2$ may be of the order of $5\%$ of $\sigma_{double}^{pD}\big|_1$. Apart from the terms representing the double parton correlation, all inputs are well known quantities. One may hence conclude that, by measuring double parton collisions in $pD$ interactions, even a reasonably limited statistics should allow obtaining a reliable indication on the size of $K(x_1,x_2)$ which, in turn, will permit a model independent estimate of the typical transverse distances between different parton pairs of the structure of the hadron.

The present discussions shows that $pD$ and, more in general, the interactions of hadrons with light nuclei represent the tool to measure the multi-parton correlations of the hadron structure, allowing not only to measure the scale factors, which quantify the rate of the different multiple parton interactions in proton-proton collisions, but also to obtain, for the first time, the additional information needed to disentangle the different correlation terms of the multiparton structure of the hadron. Although interactions of deuterons with nuclei and between polarized proton beams have been studied at RHIC, no plans are unfortunately foreseen at RHIC to study the interactions of protons with light nuclei. In spite of the limited rate of double parton collisions expected at the RHIC energy, double parton collisions have been measured by the AFS collaboration in proton-proton collisions at the CERN ISR\cite{Akesson:1986iv}, where the c.m. energy was sizably smaller. Measuremets of double parton collisions in $pD$ interactions should hence be feasible at RHIC, while the inclusion of the study of $pD$ interactions in the RHIC Physics Program would give the possibility to obtain unprecedented information on the two-body correlations between partons of the hadron structure and hence on the three dimensional parton structure of the hadron. Of course the study of double parton collisions in $pD$ interactions would be greatly facilitated at the LHC, where the much larger parton luminosity would allow to perform a detailed study of various multi-parton scattering channels in a much broader range of fractional momenta, which might represents a good motivation to support the option of running, at some stage, light nuclear beams at the LHC. 

\appendix
\section{\bf The deuteron wave function}

\vskip.25in

One may consider the coupling of the photon to the deuteron as shown in fig 6.

\begin{figure}[h]
\begin{center}
\includegraphics[width=120mm]{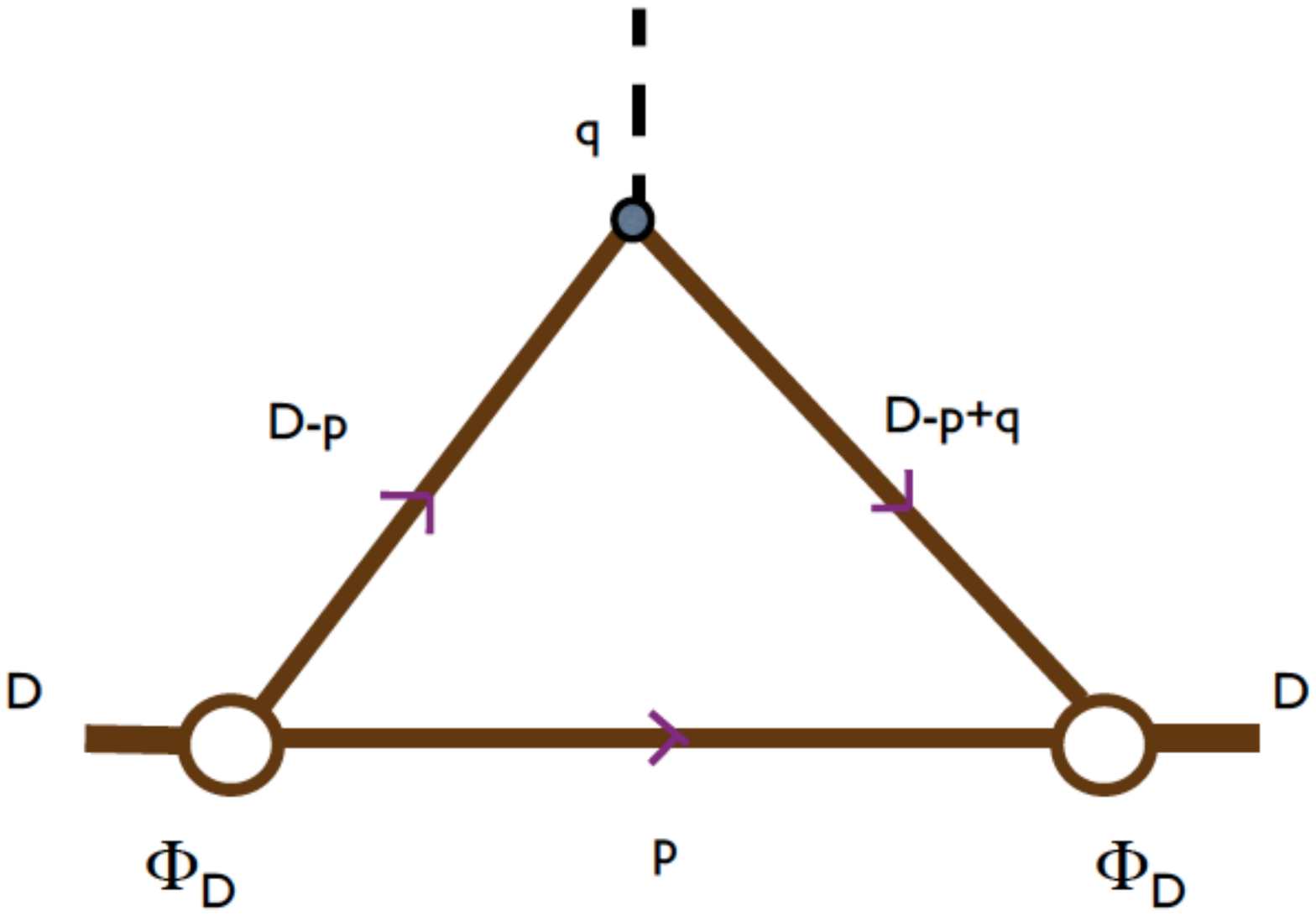}
\caption {Photon-deuteron vertex}
\label{Deuteron wave function}
\end{center}
\end{figure}

In the simplest case the vertex $\Phi_D$ is a constant and the nucleons can be treated as scalars. The expression of the diagram is

\begin{eqnarray}
i\int\frac{\Phi_D}{(D-p)^2-m^2+i\epsilon}\frac{2D_{\mu}-2p_{\mu}+q_{\mu}}{p^2-m^2+i\epsilon}\frac{\Phi_D}{(D-p+q)^2-m^2+i\epsilon}\frac{d^4p}{(2\pi)^4}=2D_{\mu}+q_{\mu}
\end{eqnarray}

\noindent
By taking the '$0$' component in the deuteron rest frame and going to the limit $q\to 0$ one must obtain:

\begin{eqnarray}
i\int\Bigl[\frac{\Phi_D}{(M_D-p_0)^2-E_p^2+i\epsilon}\Bigr]^2\frac{1}{p_0^2-E_p^2+i\epsilon}\frac{M_D-p_{0}}{M_D}\frac{dp_0d^3p}{(2\pi)^4}=1
\end{eqnarray}

\noindent
where $E_p=\sqrt({\bf p}^2+m^2)$. An equivalent expression is

\begin{eqnarray}
-\frac{i}{2M_D}\frac{\partial}{\partial M_D}\int\Bigl[\frac{\Phi_D^2}{(M_D-p_0)^2-E_p^2+i\epsilon}\Bigr]\frac{1}{p_0^2-E_p^2+i\epsilon}\frac{dp_0d^3p}{(2\pi)^4}=1
\end{eqnarray}

\noindent
One may integrate by taking the residues at $p_0=E_p-i\epsilon$ and at  $p_0=E_p+M_D-i\epsilon$

\begin{eqnarray}
\frac{-2\pi }{2M_D}&&\frac{\partial}{\partial M_D}\int\Bigl[\frac{1}{2E_p}\frac{\Phi_D^2}{(M_D-E_p)^2-E_p^2}+\frac{1}{2E_p}\frac{\Phi_D}{(M_D+E_p)^2-E_p^2}\Bigr]\frac{d^3p}{(2\pi)^4}\nonumber\\
=\frac{-1}{2M_D}&&\frac{\partial}{\partial M_D}\int\Bigl[\frac{1}{2E_p}\frac{2M_D^2\Phi_D^2}{M_D^4-(2E_pM_D)^2}\Bigr]\frac{d^3p}{(2\pi)^3}\nonumber\\
=&&\frac{-\partial}{\partial M_D^2}\int\Bigl[\frac{2\Phi_D^2}{M_D^2-4E_p^2}\Bigr]\frac{d^3p}{2E_p(2\pi)^3}=\int\Bigl[\frac{2\Phi_D^2}{(M_D^2-4E_p^2)^2}\Bigr]\frac{d^3p}{2E_p(2\pi)^3}=1
\end{eqnarray}

\noindent
one hence obtains the normalization condition

\begin{eqnarray}
\int\frac{2\Phi_D^2}{(M_D^2-4E_p^2)^2}\frac{d^3p}{2E_p(2\pi)^3}=1
\end{eqnarray}

\noindent
Notice that

\begin{eqnarray}
\frac{\Phi_D}{(D-p)^2-m^2}\Big|_{p^2=m^2}=\frac{\Phi_D}{M_D^2-4E_p^2}=\frac{\Phi_D}{-4(mB+{\bf p}^2)}
\end{eqnarray}

\noindent
where the last equation holds in the non relativistic limit and $B$ is the deuteron binding energy ($M_D=2m-B$). For constant $\Phi_D$, the last expression above coincides with the asymptotic limit at large distances of the deuteron wave function:

\begin{eqnarray}
\int\frac{e^{-\kappa r}}{r}e^{i{\bf p}\cdot{\bf r}}d{\bf r}=\frac{4\pi}{\kappa^2+{\bf p}^2}
\end{eqnarray}

\noindent
More in general one may obtain a relation between the vertex function $\Phi_D$ and the non relativistic solution of the Schr\"odinger equation by considering the solution of the Bethe-Salpeter equation in the scalar case, in ladder approximation\cite{Salpeter:1}\cite{Salpeter:2}

\begin{eqnarray}
\chi(P,p)=\frac{1}{\Bigl[\Bigl(\frac{P}{2}+p\Bigr)^2-m^2+i\epsilon\Bigr]\Bigl[\Bigl(\frac{P}{2}-p\Bigr)^2-m^2+i\epsilon\Bigr]}\int\frac{ig^2}{q^2-\mu^2}\chi(P,p+q)\frac{d^4q}{(2\pi)^4}
\end{eqnarray}

\begin{figure}[h]
\begin{center}
\includegraphics[width=120mm]{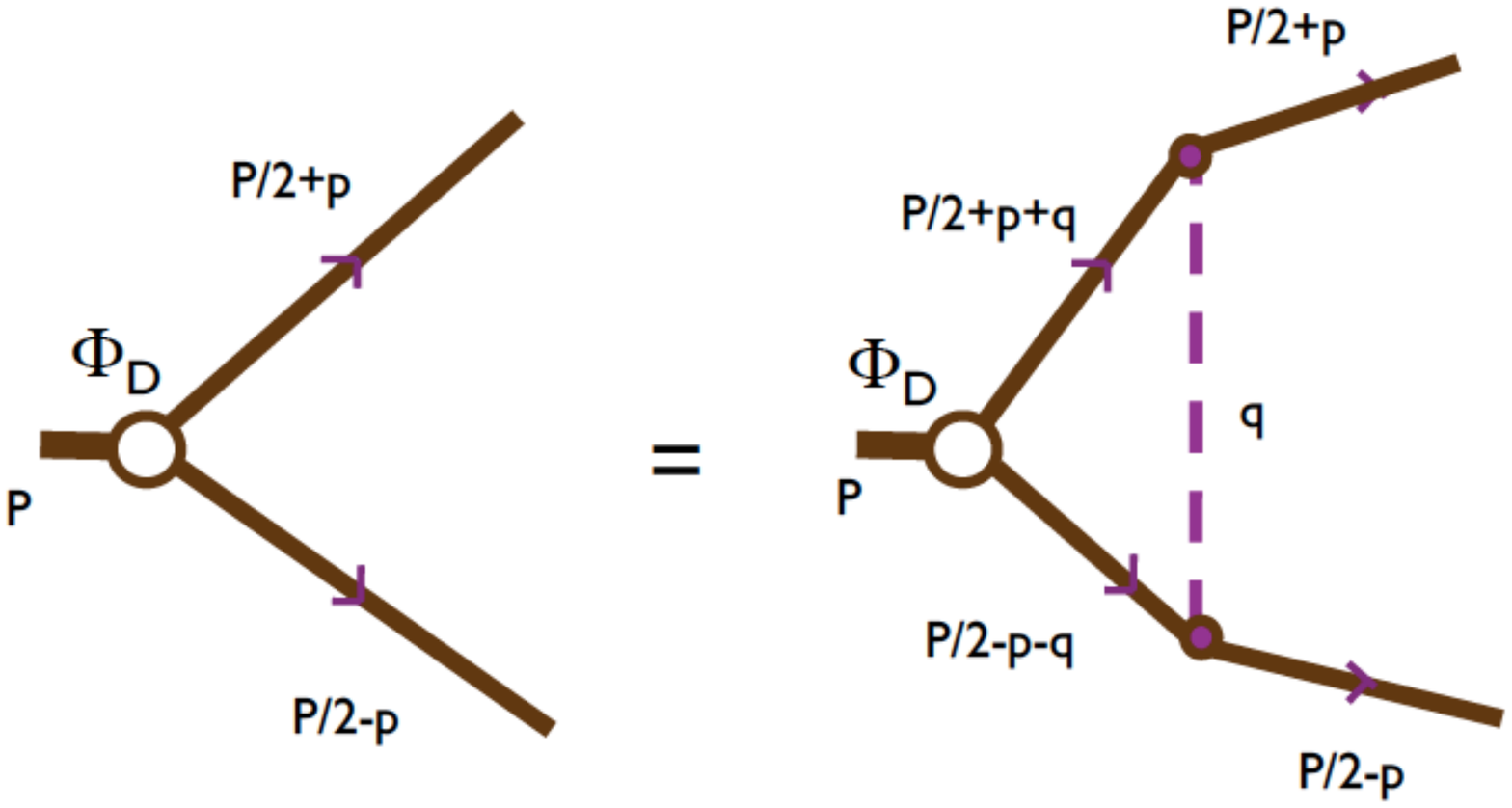}
\caption {Bethe-Salpeter equation}
\label{Bethe-Salpeter equation}
\end{center}
\end{figure}

\noindent
where $P$ is the deuteron momentum and $p$ is the nucleon momentum in the deuteron rest frame (see Fig. 7). The binding is given by the exchange of a particle of mass $\mu$ coupled to the nucleons with the constant $g$. The vertex function $\Phi_D$ is hence expressed in terms of the solution of the Bethe-Salpeter equation $\chi$ as:

\begin{eqnarray}
\Phi_D(p)=\int\frac{ig^2}{q^2-\mu^2}\chi(P,p+q)\frac{d^4q}{(2\pi)^4}
\end{eqnarray}

In the deuteron rest frame one assumes instantaneous interaction, in such a way that

\begin{eqnarray}
\frac{ig^2}{q^2-\mu^2}=\frac{-ig^2}{\bf{q}^2+\mu^2},\ {\rm and}\ \int\frac{ig^2}{q^2-\mu^2}\chi(P,p+q)\frac{d^4q}{(2\pi)^4}=\int\frac{-ig^2}{\bf{q}^2+\mu^2}\varphi_P({\bf{p+q}})\frac{d^3q}{(2\pi)^3}
\end{eqnarray}

\noindent
where

\begin{eqnarray}
\varphi_P({\bf{p+q}})\equiv\int\chi(P,p+q)\frac{dq_0}{2\pi}
\end{eqnarray}

\noindent
In the deuteron rest frame one may hence integrate the Bethe-Salpeter equation in $q_0$ as follows

\begin{eqnarray}
\int\chi(P,p)\frac{dq_0}{2\pi}&=&\int\Bigl[\Bigl(\frac{P}{2}+p\Bigr)^2-m^2+i\epsilon\Bigr]^{-1}\Bigl[\Bigl(\frac{P}{2}-p\Bigr)^2-m^2+i\epsilon\Bigr]^{-1}\frac{dq_0}{2\pi}\nonumber\\&\times&\int\frac{-ig^2}{\bf{q}^2+\mu^2}\varphi_P({\bf{p+q}})\frac{d^3q}{(2\pi)^3}
\end{eqnarray}

\noindent
which gives

\begin{eqnarray}
\varphi_P({\bf p})=\frac{1}{E_p\big(4E_p^2-M_D^2\bigr)}\int\frac{g^2}{\bf{q}^2+\mu^2}\varphi_P({\bf{p+q}})\frac{d^3q}{(2\pi)^3}
\end{eqnarray}

\noindent
In the non relativistic limit one has

\begin{eqnarray}
M_D^2=(2m-B)^2\simeq4m(m-B),\quad E_p=({\bf p}^2+m^2)^{1/2}\simeq m
\end{eqnarray}

\noindent
and one obtains

\begin{eqnarray}
\Bigl(\frac{\bf p^2}{m}+B\Bigr)\varphi_P({\bf p})=\frac{1}{4m^2}\int\frac{g^2}{\bf{q}^2+\mu^2}\varphi_P({\bf{p+q}})\frac{d^3q}{(2\pi)^3}
\end{eqnarray}

\noindent
which may be written as

\begin{eqnarray}
\frac{\bf p^2}{2m_R}\varphi_P({\bf p})+\int V({\bf q})\varphi_P({\bf p+q})\frac{d^3q}{(2\pi)^3}=E\varphi_P({\bf p})
\end{eqnarray}

\noindent
where $m_R=m/2$ is the reduced mass of the two nucleons, $E=-B$ and $V({\bf q})$ is obtained by comparing the two equations above. $\varphi_P({\bf p})$ is hence the bound state solution of the Schr\"odinger equation of the two nucleons' system. One may notice that, as in the scalar case considered here $\Phi_D$ has the dimensions of a mass, $\Psi_P$ has dimensions ${\rm mass}^{-3}$ and $\varphi_P$ ${\rm mass}^{-2}$. The bound state solution of the Schr\"odinger equation of the two nucleons' system $\varphi_P({\bf p})$ cannot hence be identified with the usual non relativistic deuteron wave function normalized to one after integration over ${\bf p}$ (which has dimensions ${\rm mass}^{-3/2}$). Obviously in the non relativistic limit, for any $\alpha$, any function $E_p^{\alpha}\varphi_P({\bf p})$ becomes a solution of the Schr\"odinger equation of the two nucleons' system. The connection with the usual non relativistic deuteron wave function may hence be obtained by determining the value of $\alpha$ through the normalization condition of the deuteron wave function:

The link between $\varphi_P$ and the vertex function $\Phi_D$ is given by

\begin{eqnarray}
\Phi_D(p)&=&\Bigl[\Bigl(\frac{P}{2}+p\Bigr)^2-m^2+i\epsilon\Bigr]\Bigl[\Bigl(\frac{P}{2}-p\Bigr)^2-m^2+i\epsilon\Bigr]\chi(P,p)
\nonumber\\&=&\int\frac{ig^2}{q^2-\mu^2}\chi(P,p+q)\frac{d^4q}{(2\pi)^4}=\int\frac{-ig^2}{\bf{q}^2+\mu^2}\varphi_P({\bf{p+q}})\frac{d^3q}{(2\pi)^3}
\nonumber\\&=&-iE_p\big(4E_p^2-M_D^2\bigr)\varphi_P({\bf p})
\end{eqnarray}

\noindent
in such a way that one may write

\begin{eqnarray}
\varphi_P({\bf{p}})=\frac{i\Phi_D(p)}{E_p\big(4E_p^2-M_D^2\bigr)}
\end{eqnarray}

\noindent
keeping into account the normalization of the Bethe-Salpeter wave function, mamely the relation

\begin{eqnarray}
\int\Bigl[\frac{2\Phi_D^2}{(M_D^2-4E_p^2)^2}\Bigr]\frac{d^3p}{2E_p(2\pi)^3}=1
\end{eqnarray}

\noindent
one can make the position

\begin{eqnarray}
\bar\varphi_P({\bf{p}}^2)=\frac{\Phi_D(p)}{\sqrt E_p\big(4E_p^2-M_D^2\bigr)}\frac{1}{(2\pi)^{3/2}}
\end{eqnarray}

\noindent
where $\bar\varphi_P({\bf{p}}^2)$ is hence normalized to 1 and, in the non relativistic limit, it is a solution of the Schr\"odinger equation of the two nucleons' system. The solution of the Bethe-Salpeter equation may hence be expressed in terms of $\bar\varphi_P({\bf{p}}^2)$ as

\begin{eqnarray}
\Psi_P(p)=\frac{\sqrt E_p\big(4E_p^2-M_D^2\bigr)\bar\varphi_P({\bf p}^2)(2\pi)^{3/2}}{\Bigl[\Bigl(\frac{P}{2}+p\Bigr)^2-m^2+i\epsilon\Bigr]\Bigl[\Bigl(\frac{P}{2}-p\Bigr)^2-m^2+i\epsilon\Bigr]}
\end{eqnarray}

\noindent
while the function $\Psi_D(p)$, defined in Eq.5, 6 is expressed in terms of $\bar\varphi_P({\bf{p}}^2)$ by the relation

\begin{eqnarray}
\frac{\Psi_D(p)}{p_-}&=&\int\frac{\Phi_D(p)}{[D-p)^2-m^2+i\epsilon][p^2-m^2+i\epsilon]}\frac{d\delta_+}{2\pi}=\frac{1}{p_-}\frac{\Phi_D}{(D-p)^2-m^2}\Big|_{p^2=m^2}\\
&=&\frac{\Phi_D}{p_-(M_D^2-4E_p^2)}=(2\pi)^{3/2}\frac{\sqrt{E_p}}{p_-}\bar\varphi_P({\bf{p}}^2)\nonumber
\end{eqnarray}

\noindent
where $\bf p$ is the three momentum in the deuteron rest frame and $E_p=\sqrt{p^2+m^2}$.

The three momentum in the deuteron center of mass frame can be expressed covariantly. In the case of interest one of the two nucleons is on shell. If, as shown in Fig.8, one puts $t=(D-p)^2$ one obtains

\begin{figure}[h]
\begin{center}
\includegraphics[width=120mm]{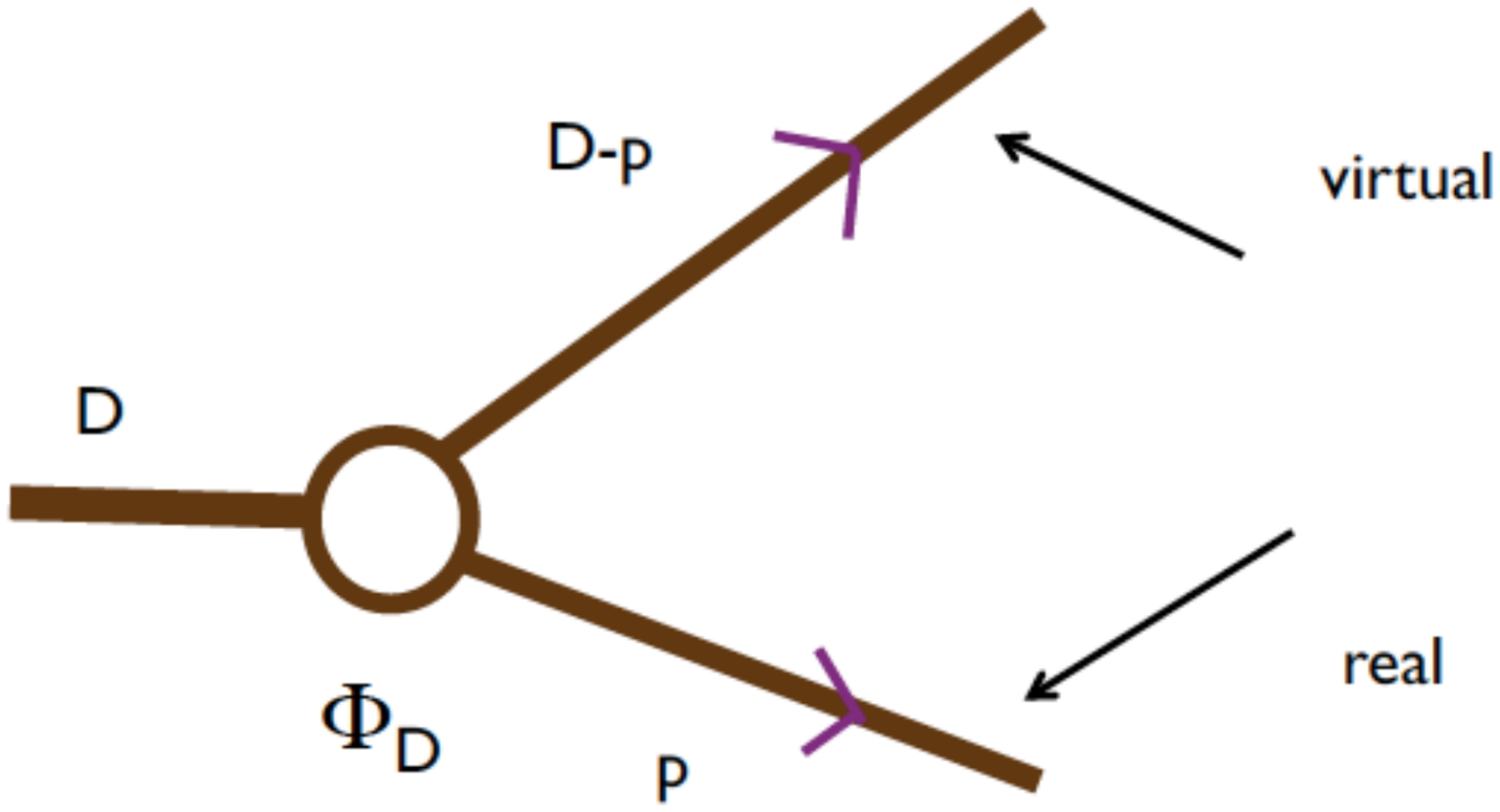}
\caption {Deuteron vertex, one nucleon is virtual and one nucleon is on shell}
\label{Deuteron splitting}
\end{center}
\end{figure}

\begin{eqnarray}
{\bf p}^2+m^2=\frac{1}{4M_D^2}(t-m^2-M_D^2)^2
\end{eqnarray}

\noindent
Introducing light cone variables, with $Z$ the momentum fraction of the virtual nucleon with respect to half of the deuteron momentum, one obtains

\begin{eqnarray}
{\bf p}^2=\frac{1}{4M_D^2}\Bigl(\frac{Z}{2}M_D^2+\frac{2}{Z}m_t^2\Bigr)^2-m^2
\end{eqnarray}

\noindent
and, for $E_p$ and $p_z$

\begin{eqnarray}
E_p=\frac{1}{2M_D}\Bigl(\frac{Z}{2}M_D^2+\frac{2}{Z}m_t^2\Bigr),\qquad p_z=\frac{1}{2M_D}\Bigl(\frac{Z}{2}M_D^2-\frac{2}{Z}m_t^2\Bigr)
\end{eqnarray}

\noindent
The function $\Psi_D(p)$ is hence finally expressed in terms of light cone variables through the non relativistic deuteron wave function $\bar\varphi_P({\bf p}^2)$ as

\begin{eqnarray}
\Psi_D(zD_-;p_t)&=&(2\pi)^{3/2}\sqrt{\frac{1}{2M_D}\Bigl(\frac{Z}{2}M_D^2+\frac{2}{Z}m_t^2\Bigr)}\nonumber\\
&\times&\bar\varphi_P\Biggl(\frac{1}{4M_D^2}\Bigl(\frac{Z}{2}M_D^2+\frac{2}{Z}m_t^2\Bigr)^2-m^2\Biggr)
\end{eqnarray}

\noindent
Notice that, since in principle $\tilde\Psi_D( Z,B)=\tilde\Psi_D( 2-Z,B)$, the expression above has to be understood as symmetrized for the exchange $Z\Leftrightarrow (2-Z)$.

\end{document}